# THERMAL EXPANSION OF GERMANIUM ISOTOPES AT LOW TEMPERATURE

R. COLELLA (A)
B. REINHART (B)
E. ALP (C)
E. E. HALLER (D)

A.) Department of Physics, Purdue University, West Lafayette IN 47907-2036
B.) Purdue Technology Center, Nesch LLC, 9800 Connecticut Drive, Crown Point IN 46307
C.) Advanced Photon Source, Argonne National Laboratory, Argonne IL 60439
D.) University of California, Berkeley, Lawrence Berkeley Laboratory, Berkeley, CA 94720

PACS number: 31.30.Gs

## ABSTRACT

The focus of this paper is the vanishing of the thermal expansion $\alpha$ in certain diamond –like semiconductors. In silicon $\alpha$ becomes zero at T = 122. K. In germanium the situation is less clear. Previous investigations show that in natural germanium a zero in $\alpha$ is reported at 15.5 K and that a second zero for T > 30 K is not observed. This paper reports on data taken on four monoisotopic Ge crystals. All plots of lattice parameter *vs* temperature shows a minimum between 40 and 50 K. The minimum is very deep for $^{70}$Ge, which also shows anomalous behaviour for the thermal expansion $\alpha$. The data are discussed with *ab initio* theoretical results obtained by Debernardi and Cardona. Some data do not support the Virtual Crystal Approximation.

The starting point of this Brief Report is the realization that the thermal expansion of Silicon changes sign at T = 122 K: it is positive for temperatures ABOVE 122 K and negative for temperatures below. The thermal expansion $\alpha$ (T) is defined as $\Delta a / a$ and depends in a complicated way on the lattice dynamics of the crystal. The thermal expansion is related to an average Grüneisen parameter G which is positive, in general, and generates a positive thermal expansion coefficient (1). The zero in $\alpha$ at 122 K is due to a NEGATIVE Grüneisen parameter of the TA phonons at the edge of the zone. The negative G at the zone edge (TA) can be explained by making use of "ab initio" calculations developed by Debernardi and Cardona (2). A good explanation can be found in a theoretical treatment (3) due to Xu, Wang, Chan and Ho. The existence of a zero in $\alpha$ (T) is an anomaly due to a combination of several factors and to peculiarities of the diamond crystal structure. Curiously enough, the Grüneisen parameter is always positive in diamond, and there is no zero in $\alpha$ (T).

What about germanium, which has the same crystal structure as diamond and silicon? The temperature dependence of the lattice constants for five isotopes of germanium ($^{70}$Ge, $^{72}$Ge, $^{73}$Ge, $^{74}$Ge, $^{76}$Ge and natural Ge) were calculated from *ab initio* electronic theory within the quasi harmonic approximation.[4] Agreement with experiment is good except for $^{73}$Ge. The theory developed by Debernardi and Cardona using perturbation theory in a density-functional framework (2) predicts that $\alpha$ (T) for silicon and germanium has TWO ZEROS, both at low temperature. Looking at a table of measured values of $\alpha$ *vs*. T in Ref. 5, Table 2, we see that $\alpha$ for Ge has a well defined zero at T = 15.5 K. The second zero should be found at higher temperature. The data for the second zero in Ref. 5 are sparse and are taken from the literature. In other words, the high temperature zero is not directly available from the data of Ref. 5. Nonetheless the authors of Ref. 5 (Smith and White) present a plot in Fig. 5 which shows the calculated variation of the average Grüneisen parameter G of Ge *vs*. T (it is called $\gamma$ in Ref. 5; it is proportional to $\alpha$ ). It shows that G is positive at high temperature; as the temperature decreases it becomes zero at about 40 K, it is then negative, it becomes zero again at T = 15.5 K, and then it is positive again up to T = 0 K. The temperature interval 15.5 - 40 K is a region in which the thermal expansion is negative. However, the experimental data, namely, the plots of lattice parameter *vs*. temperature, which are used to derive the temperature dependence of $\alpha$ and the average Grüneisen parameter G, are not given, and it is not clear that a zero in $\alpha$ has actually been observed at T > 30 K. The calculations (*ab initio*, Ref. 2) and the data show the double zero discussed for Ge, even though the results are for GaAs. However, GaAs and Ge are almost the same, in this regard.

The experiments on Ge described in Ref. 5 have been done using a "natural" Ge sample, whose atomic mass is 72.64 atomic mass units (amu). It would be of interest to determine the temperature dependence of $\alpha$ (T) *vs*. T for different isotopes, and see how they contribute to the "average" $\alpha$ values used to calculate the plot of $\gamma$ *vs*. T in Fig. 5 of Ref. 5.

For all these reasons we felt that the behavior of Germanium at temperatures below 40 K was not well characterized and that the plots of lattice parameter *vs*. T should be done with very high accuracy, using monoisotopic crystals. Four isotopically enriched bulk Ge crystals have been used in this study, with the following degrees of enrichment: 85.1 % for $^{76}$Ge, 96.8 % for $^{74}$Ge, 96.0 % for $^{73}$Ge, 96.3 % for $^{70}$Ge. These are the same crystals used in our previous experiment described in Ref. 6. One of the advantages of using bulk crystals is that they have very low dislocation densities, and therefore they are extremely perfect and yield very sharp x-ray rocking curves, a primary requirement for high resolution. More details on sample preparation are given in Refs. 6 and 7. The experimental technique (backscattering) used to measure lattice parameters with high accuracy (say: a few parts per million for $\Delta a/a$) is described in detail in Ref. 6 and will not be repeated here. Figures 1 through 4 are the experimental plots of lattice parameter *vs*. T for 4 different isotopes. Note that one increment in lattice parameter on the vertical scale corresponds to $2 \times 10^{-5}$ Å and that the temperature increment on the horizontal scale corresponds to 1 K. The good quality of the data suggest an overall accuracy of about 2 ppm (parts per million), except for Fig. 4, reporting data for $^{76}$Ge. In all these plots the smoothed profiles (black continuous line) are splines, obtained by means of polynomial expressions. The fact that the smoothed profiles are so close to the experimental points gives some confidence in the accuracy with which the lattice parameters and the thermal expansion coefficient $\alpha$ (see Fig.7) are obtained from these measurements.

Fig. 6 has been plotted using a vertical increment of which $2 \times 10^{-4}$ Å which is ten times larger than the value used in Figs. 1 to 5. The important features present in all previous plots are no longer visible. So, an accuracy of $2 \times 10^{-5}$ Å, which is a respectable value for routine work, would not reveal any special feature in the plots of $\Delta a/a$. Notice that the plots of Fig. 6 tell you very clearly that at 0 K the lattice parameters of the various isotopes are different by sizeable amounts, but they do not tell you anything about the temperature dependence of the various isotopic species between, say, 0 K and 100 K.

It is new and valuable information, not available anywhere before this work. All plots show a minimum between 40 and 50 K, which means that all isotopes exhibit a zero value in the thermal expansion coefficient. The minimum is barely visible for $^{73}$Ge and $^{Nat}$Ge, but it is prominent in the plot of $^{70}$Ge (see Fig. 1). There is no question that the isotope effect on lattice constants is much stronger for $^{70}$Ge than for the other isotopes and for $^{Nat}$Ge (M = 72.6) An important anomaly immediately strikes us: while the general rule is that isotopes with heavier nuclei have smaller lattice constants, because phonon frequencies are inversely proportional to the square roots of the nuclear masses, an important exception is immediately visible: the lattice constant of $^{73}$Ge at 0 K is LARGER than that of $^{Nat}$Ge (M = 72.64). This result casts some doubts on the validity of the so called: "Virtual Crystal Approximation", which consists in treating a crystal, which in general is a mixture of different isotopes, as a monoisotopic substance with a nuclear mass equal to the "average" value. In the case of Germanium this value would be: 72.64 amu. If we consider the thermal expansion coefficient $\alpha$, again we see that $^{70}$Ge shows a much larger effect (namely: a deeper minimum) than all other isotopes, including $^{Nat}$Ge (see Fig. 7). Note that in the plot of Fig. 7 an ordinate equal to zero corresponds to a minimum in the plots of lattice parameters vs. T. Considering the intersection of a horizontal straight line (parallel to the T axis) with ordinate = 0, it is clear that these intersections correspond to different temperatures. In other words, the minima in the plots of $^{70}$Ge, $^{73}$Ge, $^{74}$Ge, $^{76}$Ge, and $^{Nat}$Ge are observed at DIFFERENT TEMPERATURES, with the largest deviation occurring for $^{70}$Ge.

In conclusion, all Ge isotopes investigated in this work present a zero in their respective thermal expansion data as a function of temperature. Anomalous behavior is observed for $^{73}$Ge (a breaking down of the Virtual Crystal Approximation? See Ref. 8) and for $^{70}$Ge, for its large value of thermal expansion at low T.

## FIGURE CAPTIONS

Fig. 1: Lattice parameter of $^{70}$Ge vs. temperature

Fig. 2: Lattice parameter of $^{73}$Ge vs. temperature

Fig. 3: Lattice parameter of $^{74}$Ge vs. temperature

Fig. 4: Lattice parameter of $^{76}$Ge vs. temperature

Fig. 5: Lattice parameter of natural germanium vs. temperature (M = 72.64 amu)

Fig. 6 The same as for Figs. 1, 2, 3, 4, and 5, except that the scale of the ordinates is $10^{-4}$Å/division, instead of 2 x $10^{-5}$Å/division.

Let us define a quantity we can call "SENSITIVITY" (S) which is the ratio of an increment in the axis of the ordinates, typically in hundredths or thousandths of an Å, divided by a corresponding increment obtainable from the plot, typically in mm. S turns out to be equal to $10^{-6}$ for plot n. 5, and to 5.88 x $10^{-6}$ for plot n. 6. This means that S is 5.88 times greater for plot n. 6 than for plot n. 5. In other words, a given fluctuation in the counting rate $\Delta u$ will produce an excursion in the axis of the ordinates which is 5.88 times smaller in plot n. 6 when compared with plot n. 5. This difference is just the result of using different scales in the axes of the ordinates.

Fig. 7: Thermal expansion coefficient vs temperature. Calculated from plots of lattice parameter vs. temperature, after smoothing.

## REFERENCES

1. M. Cardona, Personal Communication.
2. A. Debernardi and M. Cardona, Phys. Rev. B **54**, 11305 (1996)
3. C.H. Xu, C.Z. Wang, C.T. Chan, and K.M. Ho, Phys. Rev. B **43**, 5024 (1991).
4. Y. Ma and J.S. Tse, Solid State Commun., **143**, 161 (2007).
5. T.F. Smith and G.K. White, J. Phys. C: Solid State Phys., **8** 2031 (1975).
6. M.Y. Hu, H. Sinn, A. Alatas, W. Sturhahn, E.E. Alp, H.-C. Wille, Yu.V. Shvyd'ko, J.P. Sutter, J. Bandaru, E.E. Haller, V.I. Ozhogin, S. Rodriguez, R. Colella, E. Kartheuser, and M.A. Villeret, Phys. Rev. B **67**, 113306 (2003).
7. K. Itoh, W.L. Hansen, E.E. Haller, J.W. Farmer, V.I. Ozhogin, A. Rudnev, and A. Tikhomirov, J. Mater. Res. **8**, 1341 (1993).
8. F. Widulle, J. Serrano, and M. Cardona, Phys. Rev. B **65**, 075206 (2002).

## Acknowledgments

The authors are deeply indebted to Professor Manuel Cardona for his help and explanations, in dealing with the data reported in this work.

Thanks are also due to Professor Sergio Rodriguez who suggested this research topic to R.C. and played an instrumental role in procuring the monoisotopic crystalline samples.

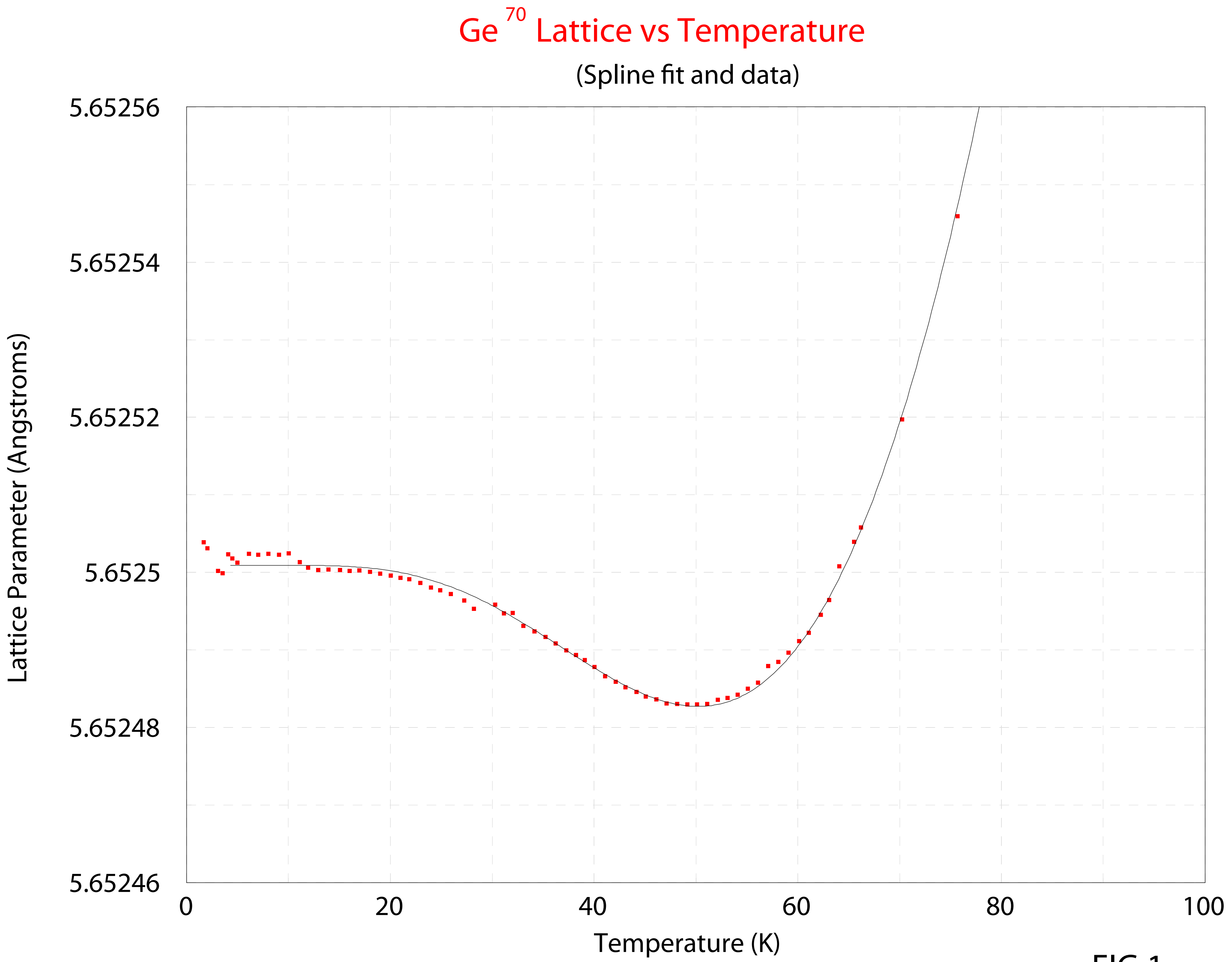

FIG 1

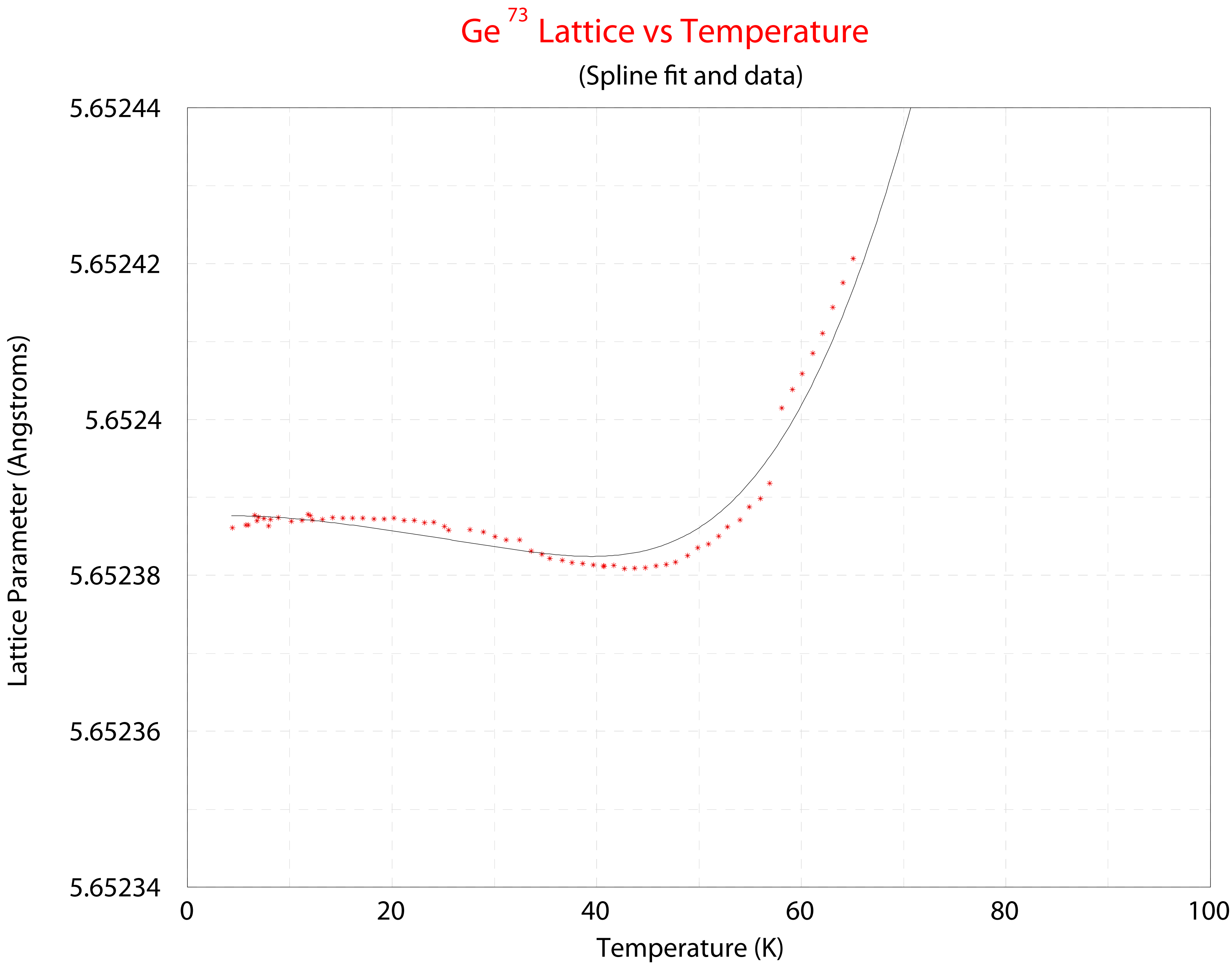

FIG 2

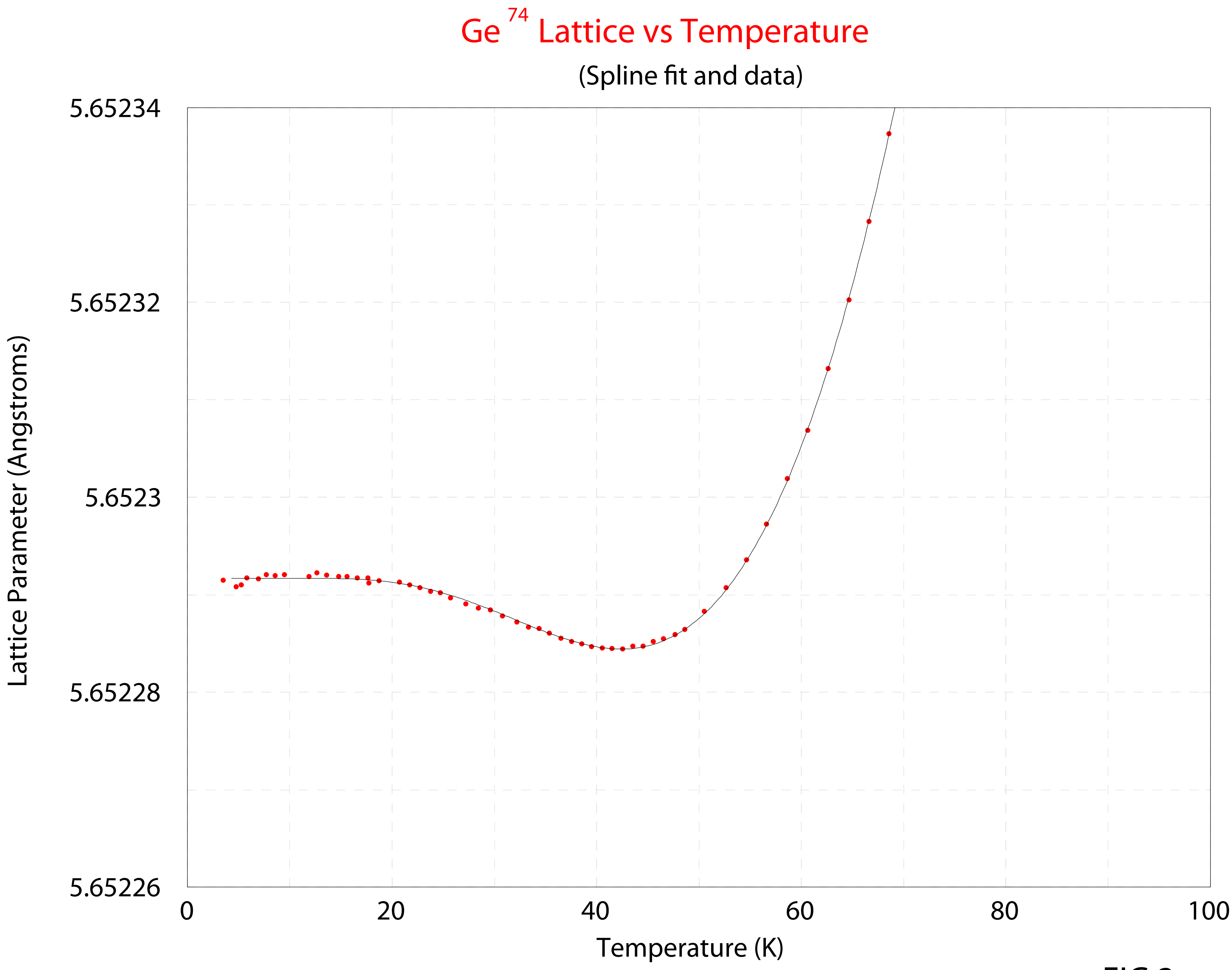

FIG 3

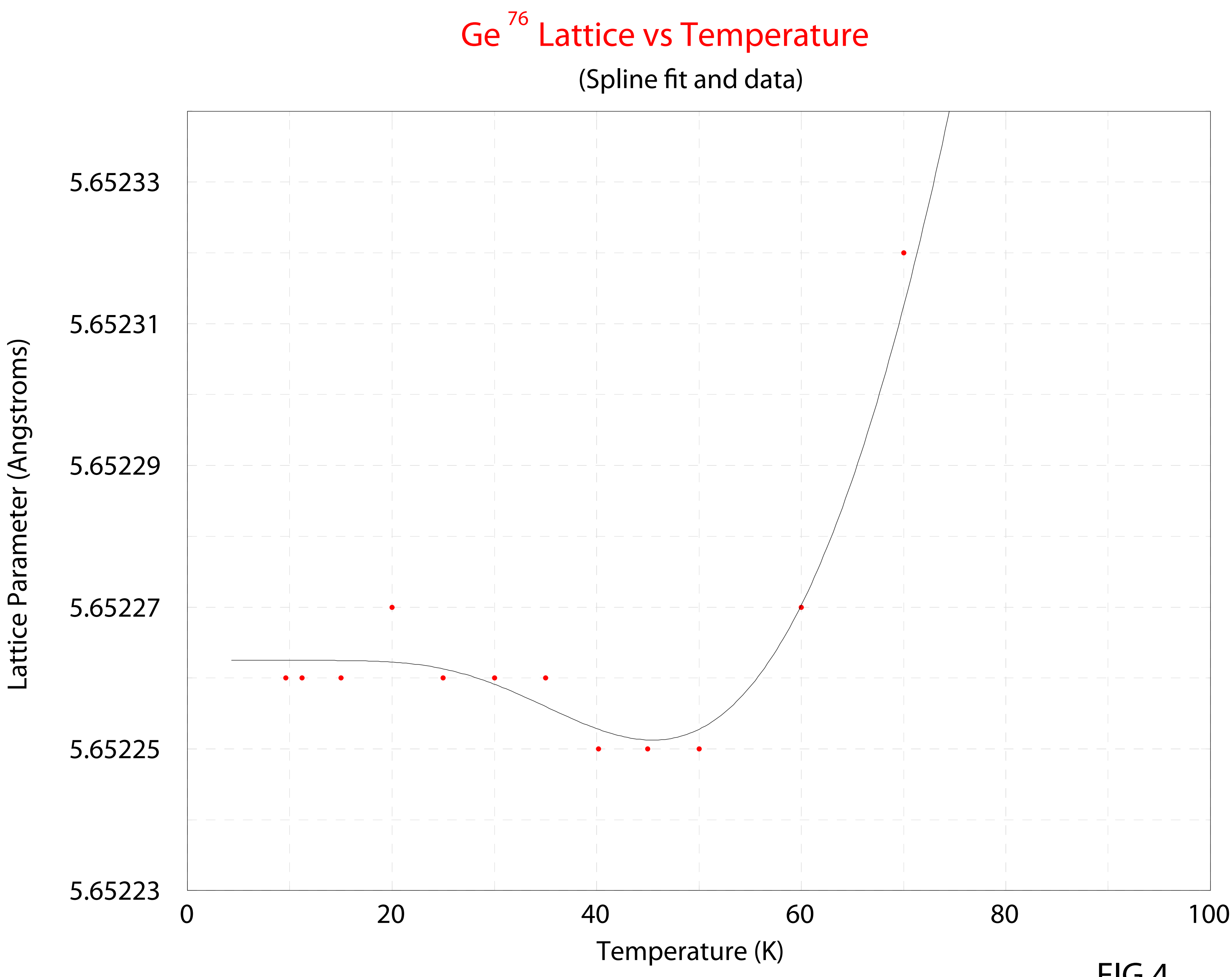

FIG 4

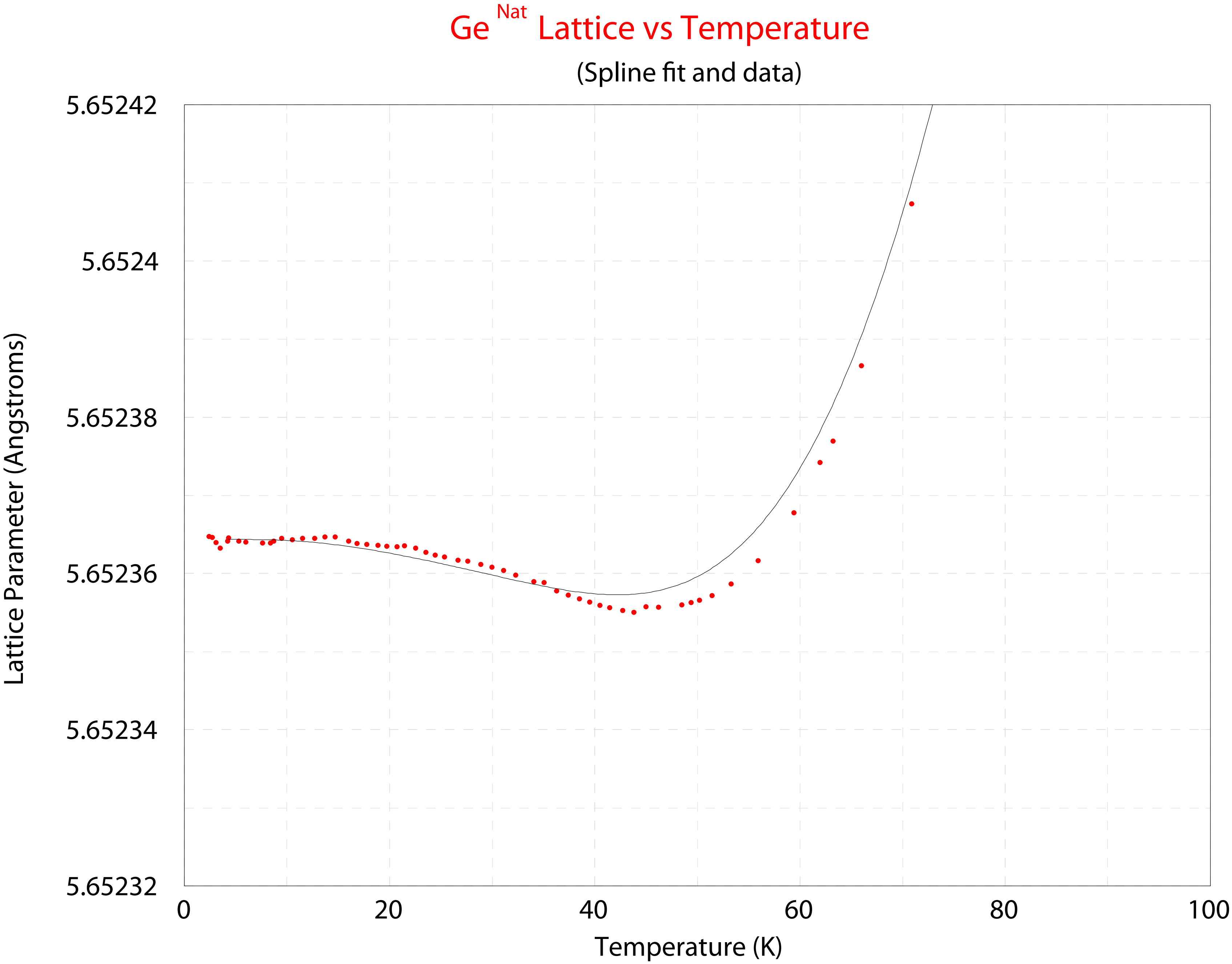

FIG 5

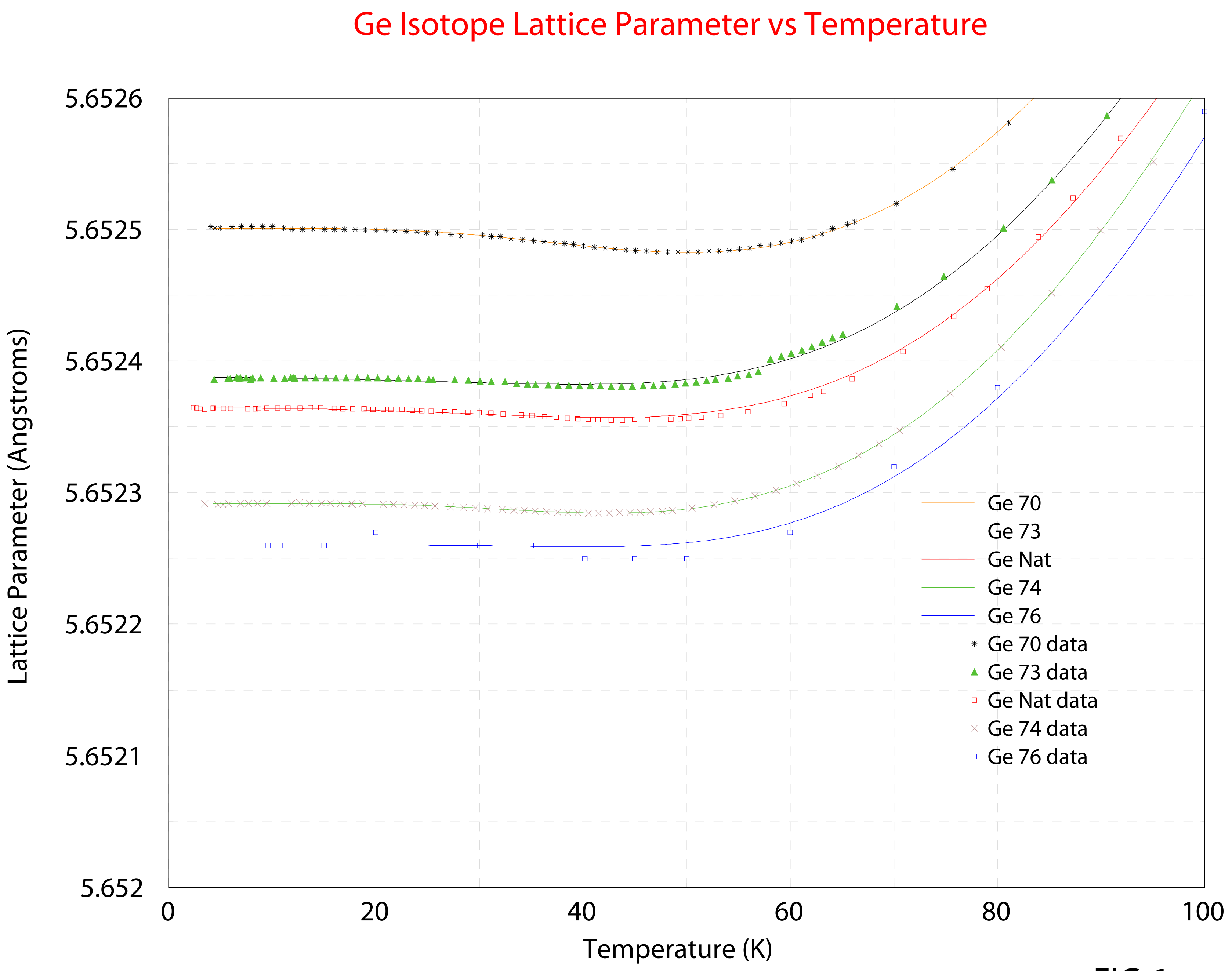

FIG 6

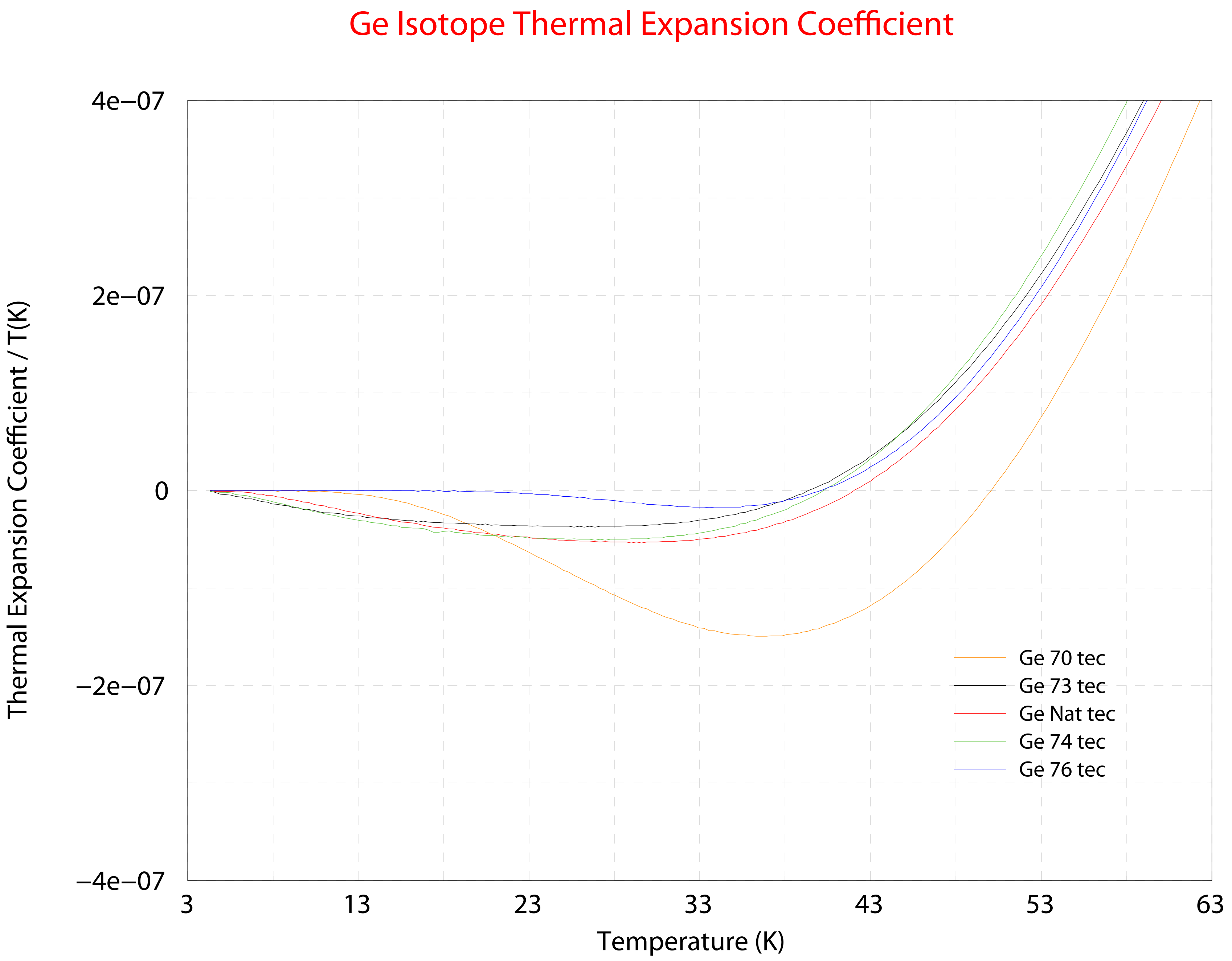

FIG 7